\documentclass[journal=jacsat,manuscript=article]{achemso}
\setkeys{acs}{keywords = true}

\usepackage{chemformula} 
\usepackage[T1]{fontenc} 
\usepackage[version=3]{mhchem} 
\usepackage{graphicx}
\usepackage{multirow}
\usepackage{rotating}
\usepackage{makecell}
\usepackage{float}

\usepackage{etoolbox}

\makeatletter
\renewcommand*{\acs@author@fnsymbol@symbol}[1]{
	\ifcase #1 *\or
	\dag\or
	\ddag\or
	\P\or
	a\or
	5\or
	6\or
	7\or
	8\or
	9\or
	10
	\fi
}

\patchcmd{\acs@address@list@auxii}
{\acs@author@fnsymbol{\acs@affil@marker@cnt}}
{\textsuperscript{\acs@author@fnsymbol{\acs@affil@marker@cnt}}}
{}{}

\patchcmd{\acs@address@list@auxii}
{{\acs@author@fnsymbol{\acs@affil@marker@cnt}\@nameuse{@altaffil@\@roman\@tempcnta}\par}}
{{\textsuperscript{\acs@author@fnsymbol{\acs@affil@marker@cnt}}\@nameuse{@altaffil@\@roman\@tempcnta}\par}}
{}{}        

\makeatother    

\author{Xiaofan Cai}
\altaffiliation{X.C., M.L. and C.C. contributed equally to this paper.}
\author{Ming Li}
\altaffiliation{X.C., M.L. and C.C. contributed equally to this paper.}
\author{Chao Chen}
\altaffiliation{X.C., M.L. and C.C. contributed equally to this paper.}
\author{Renjun Du}
\author{Zijing Guo}
\author{Ping Wang}
\author{Guodong Ma}
\author{Xinglong Wu}
\author{Zhiyuan Wang}
\author{Yaqing Han}
\author{Fuzhuo Lian}
\affiliation[NJU]
{National Laboratory of Solid-State Microstructures, School of Physics, Nanjing University, Nanjing, 210093, China. }
\alsoaffiliation[CMI]
{Chongqing 2D Materials Institute, Liangjiang New Area, Chongqing, 400714, China. }
\author{Jingkuan Xiao}
\author{Siqi Jiang}
\author{Lei Wang}
\author{Alexander S. Mayorov}
\email{mayorov@nju.edu.cn}
\author{Libo Gao}
\affiliation[NJU]
{National Laboratory of Solid-State Microstructures, School of Physics, Nanjing University, Nanjing, 210093, China. }
\author{Kostya S. Novoselov}
\affiliation[NUS]
{Institute for Functional Intelligent Materials, National University of Singapore, Building S9, 4 Science Drive 2, Singapore 117544. }
\email{kostya@nus.edu.sg}
\author{Geliang Yu}
\affiliation[NJU]
{National Laboratory of Solid-State Microstructures, School of Physics, Nanjing University, Nanjing, 210093, China. }
\email{yugeliang@nju.edu.cn}

\title{Blood Works for Graphene Production}


\begin{document}
	\newpage
		\begin{abstract}
		Blood, a ubiquitous and fundamental carbohydrate material composed of plasma, red blood cells, white blood cells, and platelets, has been playing an important role in biology, life science, history, and religious study, while graphene has garnered significant attention due to its exceptional properties and extensive range of potential applications. Achieving environmentally friendly, cost-effective growth using hybrid precursors and obtaining high-quality graphene through a straightforward CVD process has been traditionally considered mutually exclusive. This study demonstrates that we can produce high-quality graphene domains with controlled thickness through a one-step growth process at atmospheric pressure using blood as a precursor. Raman spectroscopy confirms the uniformity of the blood-grown graphene films, and observing the half-integer quantum Hall effect in the measured devices highlights its outstanding electronic properties. This unprecedented approach opens possibilities for blood application, facilitating an unconventional route in graphene growth applications.
		\end{abstract}

\section{Introduction}
		
		Because of the exceptional electronic and mechanical properties of graphene \cite{neto2009electronic}, substantial efforts have been dedicated to the chemical vapor deposition (CVD) growth of high-quality graphene using pure substances that provide single, stable carbon sources, such as methane \cite{li2011large}, ethanol \cite{guermoune2011chemical}, PMMA \cite{sun2010growth}, and various others \cite{chae2009synthesis,srivastava2010novel,xu2011production,eom2009structure}. However, the growth of graphene using compounds containing complex carbon precursors, which are both cost-effective and environmentally friendly, is constrained by multiple processing steps and the requirement for extreme conditions during growth \cite{rosmi2016synthesis,ruan2011growth,sharma2014synthesis}. Furthermore, the quality of these graphene films often falls short of expectations \cite{jalani2016defect,ray2012preparation,liu2014synthesis}. The challenge of growing high-quality graphene with precise control over the number of layers using hybrid carbon precursors remains a significant obstacle.
		
		Blood, a fluid connective tissue composed of plasma, blood cells, and platelets, has attracted people’s attention over centuries, not only because of the reverence and worship for the mysterious life, but also because of recognition as the source of life. Blood hosts a wide array of elements, including C, H, O, N, K, Ca, Na, Mg, Fe, S, Cl and so on. Virtually all animals with a circulatory system possess blood. On one hand, blood is an abundant and easily accessible resource. On the other hand, it is a complex mixture of compounds, primarily consisting of amino acids and nutrient proteins, including hydrocarbons (See Table 1).
		
		When considering the significance of using blood versus other organic materials for graphene production, it’s essential to highlight several key points. From the perspective of	abundance and sustainability, blood is a readily available and sustainable source of organic material, making it an attractive option for graphene production. What's more, graphene and graphene oxide could elicit adverse effects such as thrombogenicity and immune cell activation through interactions with blood proteins and biological membranes, which is attributed to their two-dimensional nature \cite{syama2016safety,palmieri2019graphene}. Graphene derived from blood could possess inherent biocompatibility, making it suitable for biomedical applications. The presence of biological molecules within the graphene structure could also impart specific biofunctional properties, potentially enabling targeted interactions with biological systems. Besides, utilizing blood as a precursor for graphene production will offer environmental and economic advantages, especially when repurposing waste or by-products from biomedical or pharmaceutical processes are involved.
		
		In this study, we utilize blood as a precursor and develop a one-step growth method for synthesizing a few-layer graphene on copper foil through atmospheric pressure chemical vapor deposition (APCVD). The surface-limited growth mechanism on copper \cite{li2009evolution} enables precise control of the graphene film's thickness, eliminating the need for additional post-growth contamination from the blood. Sizeable, uniform, and high-quality monolayer graphene domains are produced by our method and assessed through Raman spectroscopy and electrical transport measurements. The hydrocarbons and inorganic salts in the blood plasma play a crucial role in the formation of high-quality graphene. When heated, the water in the blood carries the solute to the surface of the copper foil. It is hydrocarbons that broaden catalytic options under different growth conditions, facilitating a dependable graphene source. Meanwhile, small sizes of inorganic salts such as NaCl are distributed across the surface of the copper foil. Large nanosheets can be produced with the assistance of the salts \cite{li20172d,shi2015direct,huang2020salt}, enabling one-step graphene growth without additional processes \cite{seo2017single}. Furthermore, we can adjust the output and proportions by modifying various growth parameters, offering a straightforward and efficient approach to producing graphene materials for diverse applications.

\section{Results and discussion}

		Graphene domains were produced through APCVD utilizing blood as a precursor on copper foil (See Supplemental Information Fig. S1), which served as the catalyst for decomposing hydrocarbons formed by hydrogen and amino acids at high temperatures, leading to the production of few-layer graphene facilitated by the low solubility of carbon in copper. As illustrated in Fig. 1(a), approximately 0.1 mL to 0.5 mL of blood was placed into a porcelain boat, filling it to one to three-quarters of its volume. Subsequently, a piece of copper foil was positioned on the boat, maintaining a slight separation from the liquid surface. The edges of the copper foil were slightly coated with blood. The porcelain boat containing the copper foil was then inserted into a quartz tube furnace. During the heating process, the water in the blood vaporized. Chemical constituents present in the blood, primarily amino acids, glucose and inorganic salts, were transported by the vapor and the constant flow of H\textsubscript{2} and catalytically decomposed into hydrocarbons on the lower side of the copper foil. After the growth process, graphene domains were observed across the entire surface of the copper foil, including the edges coated with blood.
		
		As the temperature increases, the catalytic process for the formation of graphene is speculated to be as follows: On one hand, amino acids  complex chemical reactions with hydrogen gas at high temperatures, catalyzed by copper. These reactions include the thermal decomposition of amino acids to produce carbon monoxide and carbon dioxide, which then react with hydrogen through the Fischer-Tropsch process to form hydrocarbons such as methane \cite{masters1979fischer}. On the other hand, glucose undergoes a carbonization reaction at high temperatures. The standard dehydration product of glucose, hydroxymethylfurfural, only serves as a double-bond acceptor and is not self-complementary. It may undergo a transfer due to the coordinating effect of cations on the open ketone form, leading to the formation of complex C-C bond cascades \cite{liu2014facile}. Once such oligomerization reactions occur, the products nucleate from the continuous phase, and graphene growth can ensue. The strong preference for two-dimensional growth points to an exclusive in-plane addition scheme and also suggests that salt ions stabilize the $\mathrm{sp^2}$-bonded layered carbon domains. This process leads to the formation of solvated $\mathrm{sp^2}$ aromatic carbon domains, which then grow into large-area graphene mediated by the salt ions \cite{liu2014facile}. 
		In fact, in high-temperature conditions, inorganic salts present in the blood, such as NaCl and Na\textsubscript{2}CO\textsubscript{3}, undergo melting after being transported to the copper foil. Molten salts create a fluid reaction environment, actively participating in reactions as precursors. Certain species, which may exhibit low solubility in water or other solvents, can manifest high solubility in molten salt \cite{liu2013salt}. The solubility of these precursors in molten salt significantly influences reaction rates, particle size, and morphology. Including molten salts is instrumental in controlling the nucleation and growth of graphene during CVD. For instance, the presence of Na\textsubscript{2}CO\textsubscript{3} has proven advantageous in thermal cyclodehydration and in-plane carbon reconstruction during graphene formation \cite{zhu2018general}. This influence extends to the size, density, and orientation of graphene domains on the substrate. The conversion of glucose to graphene within the molten salt system (KCl) involves the transformation of 3D $\mathrm{sp^3}$ C=X bonds (where X = C, H, or O) to 2D $\mathrm{sp^2}$ C=C bonds through the elimination of water \cite{liu2014facile,zhang2014mesoporous}. Throughout this transformation process, various structures can form within the liquid system, and carbon species can crystallize and be separated through solvent crystallization. Notably, cations play a crucial role in the synthesis of graphene in a molten salt system by assisting in the opening of keto forms of dehydration products of glucose. Additionally, cations exert a stabilizing influence on aromatic $\mathrm{sp^2}$ carbon domains, contributing to the overall efficacy of the synthesis process. If the temperature does not reach a sufficiently high level,  for example, $856\ \mathrm{^\circ C}$, which is the melting temperature of Na\textsubscript{2}CO\textsubscript{3}, the melting of the inorganic salts is inadequate. The oxygen content in the precursor became significant and the transformation of C=O bonds to C=C bonds was incomplete, resulting in the growth of graphene oxide (GO) films rather than graphene on the Cu foil (Supplemental Information Fig. S2). Due to the diverse chemical composition of blood, rather than being a single, pure substance, the lack of uniformity and limited diffusion growth occurs during the formation of graphene, as different precursor surfaces have varying diffusion rates. \cite{wu2013self,zhuang2019morphology}. After growth, all the soluble inorganic salts are easily and completely removed by washing during wet-transfer \cite{fechler2013salt,porada2015capacitive}, which in part results in the blank at the center of graphene domains shown below. Hence, a plausible hypothesis for the formation of graphene domains involves the nucleation process led by glucose under the influence of molten salt at high temperatures, followed by the growth of graphene through the formation of hydrocarbons derived from amino acids catalyzed by copper. 
		
		Fig. 1(b) and Fig. S3 depict typical graphene islands taking on the form of hexagons and snowflakes on copper foil. Selected area electron diffraction (SAED) patterns were collected to confirm the hexagonal ring structure of blood-grown monolayer graphene. Furthermore, we conducted atomic force microscopy (AFM), X-ray photoelectron spectroscopy (XPS), and energy-dispersive X-ray spectroscopy (EDX) measurements to characterize further the quality of the graphene (Supplemental Information Fig. S4-S6). Our studies unequivocally establish a simple and convenient method for growing graphene using blood as a precursor.
		
		The ability to control the thickness of graphene by adjusting the growth parameters holds significant importance for technological applications \cite{murdock2013controlling}. To assess the reliability of this growth process and the associated conditions, we investigated the growth of graphene on copper foil under different H\textsubscript{2} flow rates while maintaining a constant Ar flow rate of 250 sccm at $1050\ \mathrm{^\circ C}$ for 30 minutes, as depicted in Fig. 2(a-f). Our research demonstrates that the thickness of graphene films derived from blood can be regulated by varying the H\textsubscript{2} gas flow rate. Graphene islands are visible in all the samples, a phenomenon attributed to the complex chemical composition of blood. 
		
		In Fig. 2(g), a statistical analysis is presented, showing the relative quantity and size distribution of graphene islands across different samples. Notably, as the H\textsubscript{2} flow rate increases from 5 sccm to 20 sccm, the number of graphene islands rises and then decreases slightly at higher flow rates. The average size of the graphene islands reaches its maximum, approximately 15 $\mathrm{\mu m}$, at a flow rate of 15 sccm.
		
		Significantly, the presence of graphene islands invariably includes the existence of monolayer graphene islands, so it would be difficult to separate bilayer graphene from the products solely through variations in experimental conditions. As shown in Fig. 2(h), the proportion of monolayer graphene, which is inversely related to bilayer or trilayer graphene with lower transparency centrally located within the monolayer, exhibits an inverse relationship with the H\textsubscript{2} flow rate. Solely monolayer graphene forms when the H\textsubscript{2} flow rate is 5 sccm, whereas more than 50\% of the graphene islands become bilayer at the H\textsubscript{2} flow rate of 30 sccm in 30 minutes. We conducted Raman spectroscopy to investigate the properties of monolayer graphene grown under different H\textsubscript{2} flow conditions, as shown in Fig. 2(i). Both the G peak and 2D peak are present under all the examined conditions. The ratios of I(2D)/I(G) range from 1.01 to 2.09, indicating the successful growth and high quality of monolayer graphene, confirming the simplicity and robustness of the method. The intensity of the D peak increases as the H\textsubscript{2} flow rate rises from 5 sccm to 30 sccm, except for at 10 sccm and 15 sccm. Redshifts are observed in the G peak and 2D peak of monolayer graphene with 30 sccm H\textsubscript{2}, signifying its hole-doping characteristic due to water's presence during the wet transfer procedure.
		
		The impact of H\textsubscript{2} can be explained as follows. It has been established that hydrogen plays a dual role during the CVD process for graphene synthesis. Firstly, it facilitates the catalytic decomposition of hydrocarbon gases, which generates more active carbon species on the exposed metal surface. Secondly, hydrogen can also etch graphene \cite{zhang2014role,vlassiouk2011role,losurdo2011graphene}.
		
		The dehydrogenation of hydrocarbons in the blood produces active carbon species ($\mathrm{C_xH_yO_z}$). When hydrogen pressure is low, only a limited number of active carbon species are generated, and the graphene edges are not effectively passivated by hydrogen. This causes the graphene to adhere tightly to the catalyst surface, preventing any active carbon species from diffusing into the layers beneath the surface\cite{zhang2014role}. Consequently, only monolayer graphene is formed, and its size remains restricted. With higher hydrogen pressures, a greater number of carbon species are produced, and the size of graphene islands reaches its maximum. This occurs due to a balance between forming active carbon species and etching graphene by hydrogen. However, when the hydrogen pressure increases, the generation of active carbon species becomes saturated, and the etching effect of hydrogen becomes dominant. This leads to the termination of the graphene edges, favoring the diffusion of active carbon species into the layers beneath the surface. As a result, the blood-grown graphene film becomes thicker, while its lateral size appears smaller.
		
		In addition, hydrocarbons broaden catalytic options in the CVD production of graphene. The use of hydrocarbons expands the range of available precursors, each with its own unique chemical and physical properties. On one hand, Hydrocarbons encompass a diverse range of structural motifs, including linear, cyclic, and aromatic carbon frameworks. In our experiments, the hydrocarbons primarily consist of alkanes (such as methane) catalyzed by amino acids and aromatic carbon frameworks inherent in aromatic amino acids, including tyrosine, phenylalanine, and tryptophan. This structural diversity offers a broad spectrum of chemical functionalities and bonding configurations, expanding the catalytic options for tailoring the growth mechanisms and morphologies of graphene during CVD. On the other hand, hydrocarbons exhibit a wide range of reactivities, which can be leveraged to modulate the growth kinetics and quality of graphene during CVD. Selecting specific hydrocarbons with different reactivity profiles makes it possible to finely tune the CVD process, influencing factors such as growth rate, nucleation density, and crystallinity of the graphene produced. This tunability provides additional catalytic options for controlling the growth and properties of graphene. 
		
		Moreover, it is worth noting that the N element does not participate in the formation of graphene. Instead, under the conditions of high growth temperature and ample hydrogen supply, the N elements present in the blood preferentially form ammonia and volatize, rather than incorporating into the graphene lattice as N doping. Evidence supporting the absence of N in the produced graphene is provided by both XPS and EDS analyses, as shown in Fig. S5 and S6. The XPS spectra fail to reveal any N 1s peak within the energy range of 397 to 412 eV, as well as the absence of N-C=O or N-C-O peaks in the C 1s spectrum and N-O or C-N peaks in the O 1s spectrum. Similarly, the EDS analysis does not detect the presence of the N K$\alpha$ peak centered around 392 eV. These observations collectively reinforce the conclusion that the generated graphene is free of N doping.
		 
		Fig. 3(a) illustrates monolayer, bilayer, and trilayer graphene domains grown using our method. We conducted Raman spectroscopy measurements to determine the thickness and stacking order and assess the graphene films' quality and uniformity. Fig. 3(b) displays Raman spectra obtained from the spots marked with correspondingly colored circles in Fig. 3(a). Typically, in all Raman spectra, characteristic peaks of the G band and 2D band are observed near 1580 $\mathrm{cm^{-1}}$ and 2680 $\mathrm{cm^{-1}}$, respectively. The low intensity of the D peaks in the samples indicates high quality with few defects across the covered surface, which is further evidenced by Raman mapping of the D peak in Fig. 3(c).
		
		For the monolayer graphene, indicated by the black circle, the Raman spectrum features a narrow and symmetric 2D peak with a ratio of I(2D)/I(G) of 2.14 and a full width at half-maximum (FWHM) of 45 $\mathrm{cm^{-1}}$, confirming the high crystalline quality of the monolayer graphene. In the case of the red circle, a broader 2D peak (FWHM of 61 $\mathrm{cm^{-1}}$) with a ratio of I(2D)/I(G) close to 1 is evident, in line with the expected characteristics of AB-stacked bilayer graphene \cite{ferrari2006raman,graf2007spatially,robertson2011hexagonal}. As for the graphene denoted by the blue circles at the center, the I(2D)/I(G) ratio is approximately 0.5, with the FWHM of the 2D peak around 70 $\mathrm{cm^{-1}}$. This closely resembles the features of exfoliated trilayer graphene with ABA stacking rather than ABC stacking, as indicated by the absence of a sharp peak and reduced asymmetry in the 2D peaks in ABA compared to ABC stacking \cite{lui2011imaging}. The 2D peaks of bilayer and trilayer graphene can be well-fitted with 4 and 6 Lorentzians, as shown in Fig. S7. 
		
		Raman maps of the D peak, G peak, 2D peak, and FWHM of the 2D peak are shown in Fig. 3(e-f). Various thicknesses of graphene can be easily distinguished in the G and 2D FWHM maps, each represented by distinctive colors with high contrast. The uniformity of the Raman maps across different thickness regions underscores the high quality and exceptional homogeneity of the graphene domains.
		
		We also examined the electrical properties of graphene synthesized from a blood precursor by conducting transport measurements using a standard four-probe method. The graphene grown on copper foil was transferred onto silicon wafers using ferric chloride or ammonium persulphate solution as an etchant in a wet-transfer method. Standard electron-beam lithography was employed to create graphene Hall bars on conventional 300 nm $\mathrm{SiO_2/Si}$ substrates. In Fig. 4(a), one can see optical images of a device, including a close-up of the graphene film before it was transferred.
		
		Fig. 4(b) displays the four-terminal resistance measured as a function of back-gate voltage at $T$ = 1.6 K. The distinct Dirac peak shifted to $V_g$ = 15 V is clearly observable, with a Hall mobility of approximately 1900 $\mathrm{cm^2\ V^{-1}\ s^{-1}}$. This allows observing the Quantum Hall effect at 1.6 K under a magnetic field of 4.5 T. The shift of the Dirac point towards positive gate voltages indicates that the transferred graphene is weakly p-type doped. While the presence of water in the blood and residual oxygen impurities from the catalyst of amino acids impose some limitations on achieving ultra-high carrier mobilities, the plateaus at filling factors $\upsilon$ = 2, 6, etc., in $R_{xy}$, can be well fitted. This evidence supports the successful growth of high-quality monolayer graphene. 
		
		Overall, the use of blood as a precursor for the one-step production of graphene exhibits significant advantages in terms of product quality, process simplicity, and cost-efficiency. In terms of product quality, the mobility of the graphene produced using blood as a precursor is much higher than that of graphene grown from liquid \cite{guermoune2011chemical}, solid precursors \cite{sun2010growth}, or mixtures \cite{seo2017single} (typically below 500 $\mathrm{cm^2\ V^{-1}\ s^{-1}}$). In terms of process method, the one-step synthesis avoids the cumbersome secondary gas/liquid injection process compared to gas and liquid precursors, making the operation more straightforward and convenient. Additionally, the cost of blood is significantly lower than other precursors, making this method cost-effective and environmentally friendly while maintaining product quality.

\section{Conclusion}

		In conclusion, we have successfully developed a straightforward and highly efficient method for producing high-quality graphene utilizing blood as a precursor. The formation of high-quality graphene is inextricably related to the efficient catalysis of numerous hydrocarbons present in blood plasma, facilitated by molten inorganic salts. The control of area coverage, ranging from monolayer to AB-stacked bilayer and ABA-stacked trilayer graphene, is achievable by adjusting the H\textsubscript{2} flow rate. The growth temperature plays a significant role in determining the chemical composition of the resulting film, with options ranging from pure carbon (graphene) to a larger quantity of graphene oxide (GO). Raman mapping and transport measurements consistently affirm the uniformity and homogeneity of blood-grown graphene via the APCVD method. Notably, it exhibits electronic properties on par with devices fabricated using graphene from pure hydrocarbon gases.
		
		The novel CVD method we have developed offers several key advantages. It is remarkably straightforward, requiring only the precise handling of the precursor, without the need for complex gas injection systems used with liquid precursors. This flexibility stems from the abundant chemical components present in the blood. Furthermore, blood is a readily available and cost-effective material, making this method a significant step towards a simplified and versatile process for synthesizing high-quality graphene using compounds with complex carbon precursors. What's more, taking into account that one liter of blood contains approximately 20 g of carbon, and with graphene’s specific surface area being 2630 $\mathrm{m^2/g}$ , and assuming a material conversion rate of 0.5\% for CVD graphene production, one liter of blood using our method could yield approximately 250 $\mathrm{m^2}$ of graphene. These advancements pave the way for cost-effective and commercially viable graphene production, opening up new application opportunities.
                                          
\section{Experimental Methods}
		\textbf{\quad\ Sample Preparation.} The copper foil (Alfa Aesar, item No. 046986) was successively immersed in alcohol, acetone, and isopropanol solutions and subjected to ultrasonic cleaning for ten minutes to eliminate organic contaminants. Fresh porcine blood, which closely approximates the chemical composition of human blood\cite{sondeen2013comparison}, was obtained from the market and then centrifuged at 5000 rpm for three minutes. The resulting supernatant, primarily composed of water, proteins, electrolytes, and so on, was collected and stored at temperatures below $4\ \mathrm{^\circ C}$ until needed. We also tried plasma from porcine blood (Sigma-Aldrich, item P2891). The freshness of the blood has minimal impact on the quality of the CVD-grown graphene. 
		
		\textbf{AFM, Raman, SEM, EDX, XPS, and TEM Measurements.}
		AFM measurements are performed with a Bruker Dimension Fastscan
		system at tapping mode. Room temperature Raman scattering was performed with a WITec/alpha 300R confocal microscope or a Horiba T64000 Raman spectrometer with an 1800 gr/mm grating using a 532 nm laser under ambient conditions. Laser power was maintained below 1 mW to prevent damage or heating. The G and 2D peaks in the Raman spectra were fitted with Lorentzian functions. The microstructures and composition of the CVD-grown graphene were characterized by scanning electron microscopy (SEM, Carl Zeiss, GeminiSEM 500) and energy dispersive spectrometer (EDS, Oxford Instruments, Ultim Extreme). The XPS analyses were carried out on a PHI 5000 VersaProbe spectrometer using a monochromatic Al K(alpha) X-ray source. The selected-area electron diffraction (SAED) measurements of the graphene were performed to evaluate the structure with an operating voltage of 200 kV on a TEM (Tecnai F20). The CVD-grown graphene samples were transferred onto the copper grid via the Ferric chloride (FeCl3) solution etching-assisted transfer method.
		
		\textbf{Transport Measurements.}
		The device was measured in an Oxford Instruments TeslatronPT cryogen-free superconducting magnet system equipped with an Oxford Instruments insert (1.6 K, 14 T), with the magnetic field applied perpendicular to the plane of the heterostructure. A Stanford Research Systems SR830 lock-in amplifier was employed to apply an AC bias current using a 100 MOhm bias resistor at 13.333 Hz. Keithley 2400 SourceMeters were used to apply voltages to the gates. 
		
\section{Acknowledgments}
		KSN is grateful to the Ministry of Education, Singapore (Research Centre of Excellence award to the Institute for Functional Intelligent Materials, I-FIM, project No. EDUNC-33-18-279-V12) and to the Royal Society (UK, grant number RSRP\textbackslash R\textbackslash 190000) for support. Geliang Yu acknowledges Financial support from the National Key R$\&$D Program of China (Nos. 2022YFA120470, 2021YFA1400400), the National Natural Science Foundation of China (Nos. 12004173, 11974169), and the Fundamental Research Funds for the Central Universities (Nos. 020414380087, 020414913201). And the authors would like to thank Mono Technology Co, Ltd (http://www.wuximono.com) for the CVD growth system, which is used for graphene and 2D materials production.
		
\section{Author contributions}
		Conceptualization: K.S.N. Methodology: G.Y., X.C., M.L., C.C. CVD: X.C., M.L. Sample fabrication: M.L., C.C., P.W., G.M. Raman: X.C., Z.G., X.W., L.G. SAED: C.C. AFM: X.C., F.L., L.G. XPS: M. L. EDS: Z.W., Y.H., C.C Electrical transport: X.C., A.S.M., J.X., S.J., F.L. Funding acquisition: K.S.N., G.Y., R.D. Discussion: X.C., M.L., C.C. Visualization: X.C. Supervision: K.S.N., G.Y., L.W. Writing-original draft: X.C., M.L. Writing-review and editing: X.C., M.L., C.C., R.D., L.W., A.S.M., L.G., G.Y., K.S.N.
		
\section{Competing interests}
		The authors declare no competing interests.

\section{Additional information}
		Correspondence and requests for materials should be addressed to Alexander S. Mayorov, Kostya S. Novoselov or Geliang Yu.

\section{Supporting Information}
		\textbf{Supporting Information Available:} CVD growth of graphene, CVD growth of graphene and graphene oxide (GO) at different temperatures, SEM images of the monolayer graphene, AFM images of fresh blood-grown graphene, XPS spectrum of the blood-grown graphene, EDX investigations on CVD blood-grown graphene, Lorentz fitting of the 2D peaks of the bilayer and trilayer graphene, Impact of water on electronic properties of the graphene
\bibliography{Ref}

\newpage
\begin{sidewaystable}[]
	\caption{Components of blood}
	\begin{tabular}{ccccc} 
		\hline
		Components of blood      & Compound                                & Example/Comment                         & Elements                                 & Content                     \\ \hline
		\multirow{15}{*}{Plasma} & Water                                   & $\mathrm{H_2O}$                                     & H, O                                     & 90\% - 92\% (in plasma)     \\
		& \multirow{3}{*}{Proteins}               & Albumins                                & C, H, O, N, S, etc.                      & 35 - 51 g/L                 \\
		&                                         & Globulins                               & C, H, O, N, S, etc.                      & 20 - 40 g/L                 \\
		&                                         & Fibrinogen                              & C, H, O, N, S, etc.                      & 2 - 4 g/L                   \\
		& \multirow{2}{*}{Amino acids}            & Alanine                                 & C, H, O, N                               &                             \\
		&                                         & Glutamine                               & C, H, O, N                               &                             \\
		& Electrolytes                            & \makecell{$\mathrm{Na^+,\ Ca^{2+},\ Mg^{2+},}$ \\ $\mathrm{HCO_3^-\ ,\ Cl^-}$}                 & \makecell{$\mathrm{K,\ Ca,\ Na,\ Mg,\ Zn,\ Fe,}$ \\ $\mathrm{C,\ H,\ O,\ N,\ P,\ S,\ Cl,\ etc.}$} &                             \\
		& Gases                                   & $\mathrm{O_2,\ CO_2}$                                 & C, O                                     & 1\% - 2\%                   \\
		& Nutrients                               & \makecell{Building blocks of \\ proteins and glucose} & C, H, O, N, S, etc.                      &                             \\
		& \multirow{3}{*}{\makecell{Nitrogenous \\ waste}}      & Urea                                    & C, H, O, N                               & 3.2 - 7.1 $\mathrm{mmol/L}$           \\
		&                                         & Creatinine                              & C, H, O, N                               & 53 - 140 $\mathrm{\mu mol/L}$              \\
		&                                         & Uric Acid                               & C, H, O, N                               & 208 - 428 $\mathrm{\mu mol/L}$            \\
		& \multirow{3}{*}{\makecell{None-nitrogenous \\ waste}} & Glucose                                 & C, H, O                                  & 3.9 - 7.8 $\mathrm{mmol/L}$            \\
		&                                         & Amino acids                             & C, H, O                                  &                             \\
		&                                         & Lipids                                  & C, H, O                                  & \textless{}1\%              \\
		Red blood cells          & Cells                                   & Hemoglobin                              & C, H, O, N, S, Fe, etc.                  & 5 million - 5.5 million/$\mathrm{mm^3}$ \\
		White blood cells        & Cells                                   & \makecell{Neutrophils, eosinophils \\ and basophils}  & C, H, O, N, S, etc.                      & 6000 - 8000/$\mathrm{mm^3}$             \\
		Platelets                & Cell fragments                          & phospholipids                           & C, H, O, N, S, etc.                      & 150000 - 350000/$\mathrm{mm^3}$        \\ \hline
	\end{tabular}
\end{sidewaystable}

\clearpage
\graphicspath{{Figures}}
\begin{figure}[H] 
	\centering
	\includegraphics[width=1\textwidth]{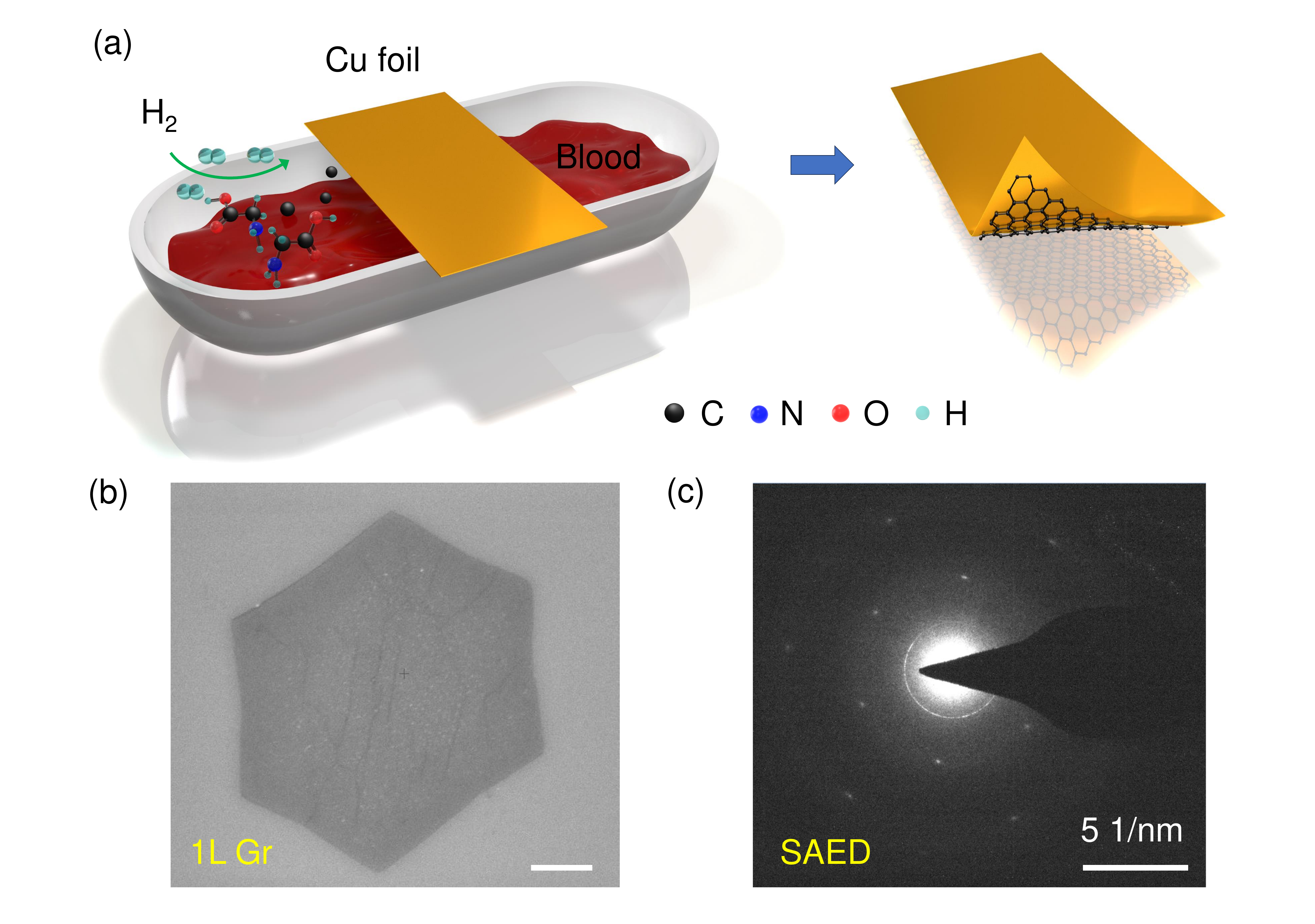}
\end{figure}
\textbf{Fig 1. CVD-grown Graphene Films Synthesized from a Blood Precursor.} (a) Schematic representation of the process for growing graphene on copper foils using blood as a precursor. (b) Scanning electron microscope (SEM) images of monolayer graphene islands with a hexagonal shape. Scale bar, 1 $\mathrm{\mu m}$. (c) The selected area electron diffraction (SAED) pattern of the blood-grown monolayer graphene distinctly reveals a hexagonal lattice.

\newpage
\graphicspath{{Figures}}
\begin{figure}[H] 
	\centering
	\includegraphics[width=1\textwidth]{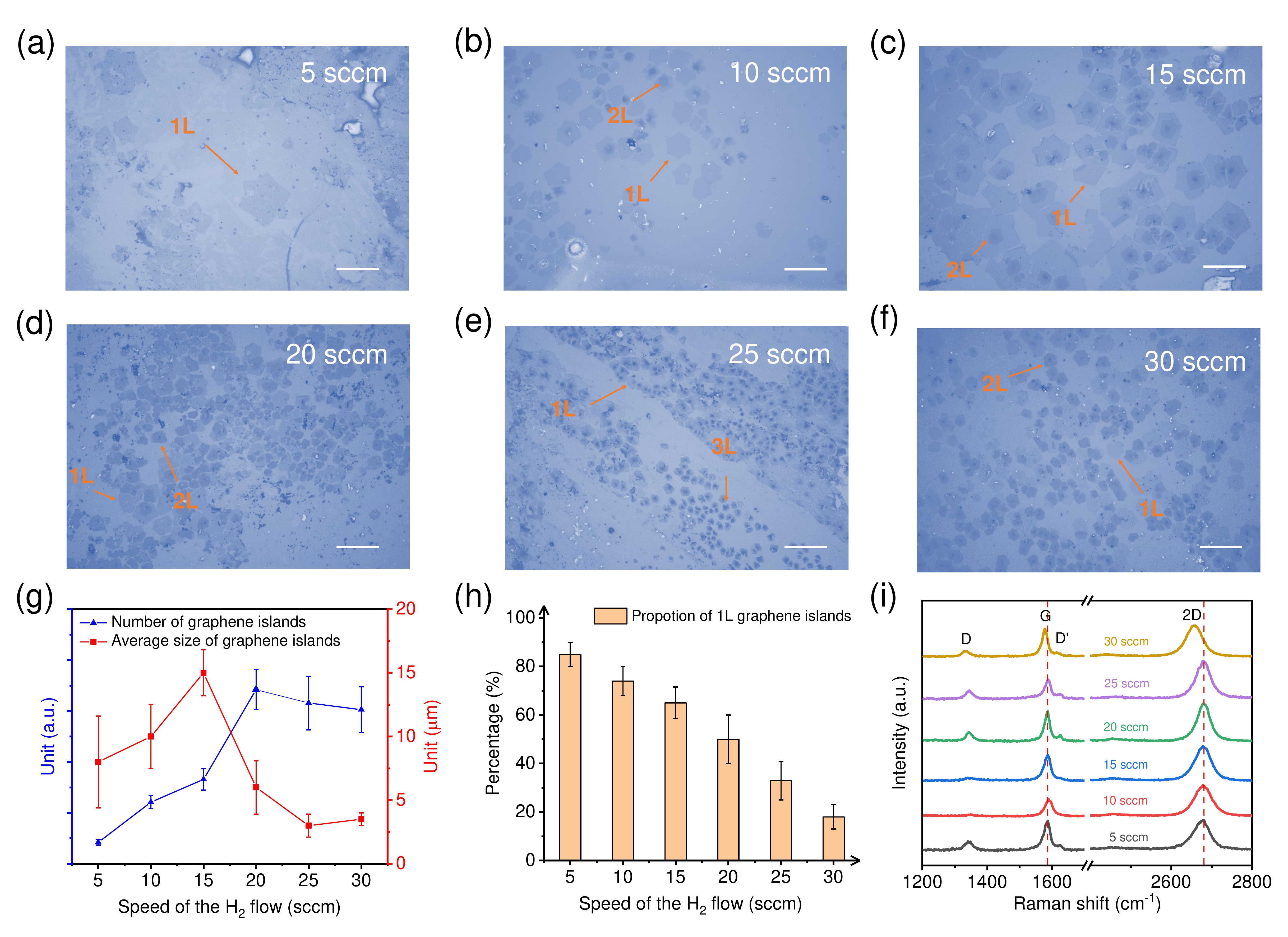}
\end{figure}
\textbf{Fig 2. Robust growing conditions.} Optical images of blood-grown graphene at different H\textsubscript{2} flow rates transferred onto $\mathrm{SiO_2/Si}$ substrates: (a) 5 sccm, (b) 10 sccm, (c) 15 sccm, (d) 20 sccm, (e) 25 sccm, (f) 30 sccm. Arrows indicate varying graphene thicknesses. Scale bars, 20 $\mathrm{\mu m}$. (g) Variations in graphene islands' number and average size with H\textsubscript{2} flow rate. The average size of graphene islands peaks at approximately 15 $\mathrm{\mu m}$ due to the combined effects of catalytic hydrocarbon decomposition and H\textsubscript{2} etching. Error bars represent the standard deviation in the size and quantity of graphene islands measured across different samples. (h) The proportion of monolayer graphene islands monotonically decreases with increasing H\textsubscript{2}\textbf{} flow rate. Error bars represent the standard deviation in the quantity of graphene islands measured across different samples. (i) Raman spectra of monolayer graphene in (a-f). Vertical dashed lines indicate the positions of the G band and 2D band at 1580 $\mathrm{cm^{-1}}$ and 2680 $\mathrm{cm^{-1}}$, respectively.

\newpage
\graphicspath{{Figures}}
\begin{figure}[H] 
	\centering
	\includegraphics[width=1\textwidth]{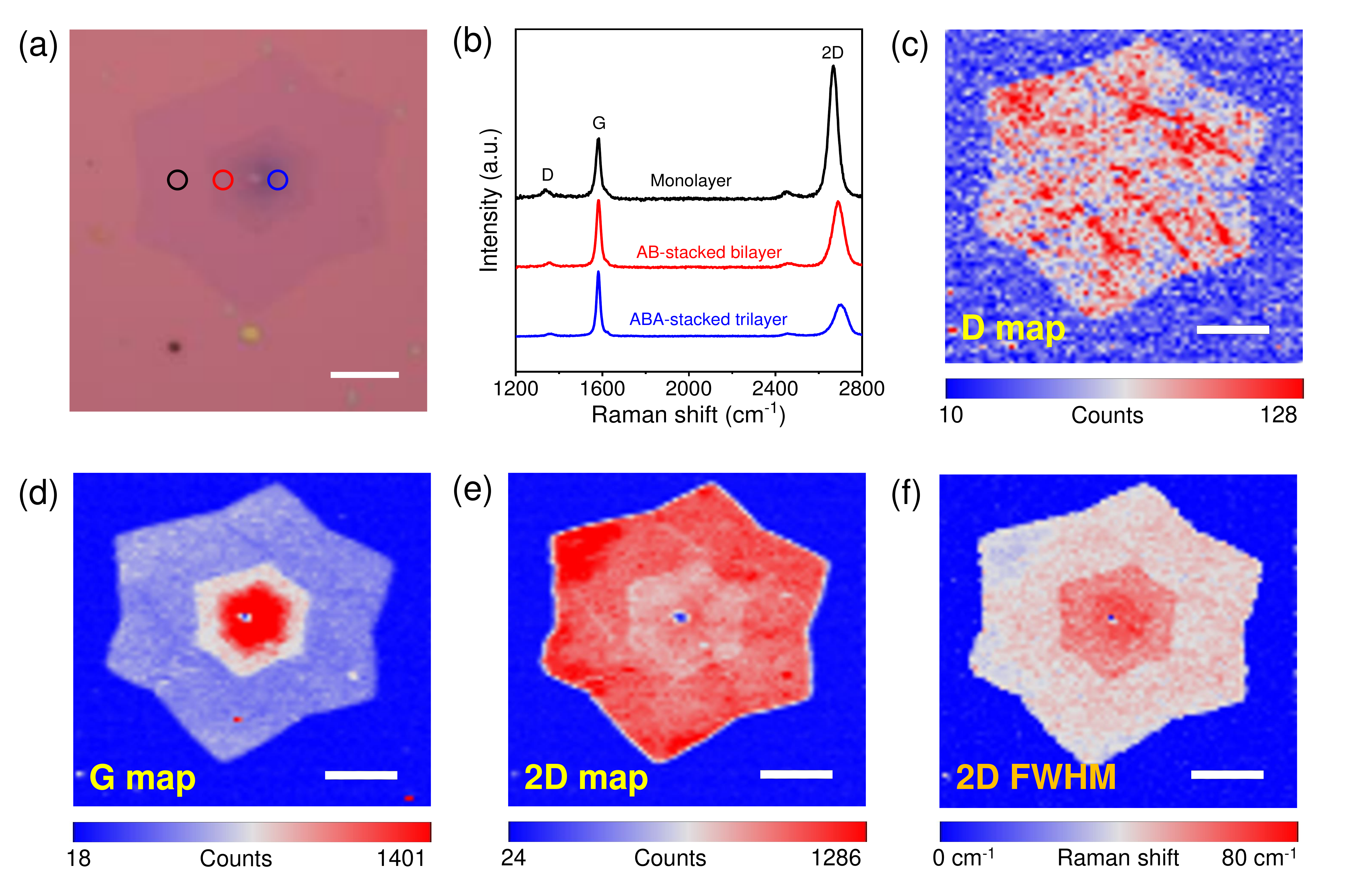}
\end{figure}
\textbf{Fig 3. Raman Measurements for Monolayer, Bilayer, and Trilayer Graphene.} (a) Optical microscope images of a graphene domain. (b) Corresponding Raman spectrum obtained from different colored circles marked in (a). All Raman spectra are normalized for clarity (a.u., arbitrary units). (c-e) Raman maps showing the D (1350 $\mathrm{cm^{-1}}$), G (1580 $\mathrm{cm^{-1}}$), and 2D (2680 $\mathrm{cm^{-1}}$) bands, respectively. (f) Raman map of the Full Width at Half Maximum (FWHM) of the 2D band for the graphene domain in (a). Scale bars, 5 $\mathrm{\mu m}$.

\newpage
\graphicspath{{Figures}}
\begin{figure}[H] 
	\centering
	\includegraphics[width=1\textwidth]{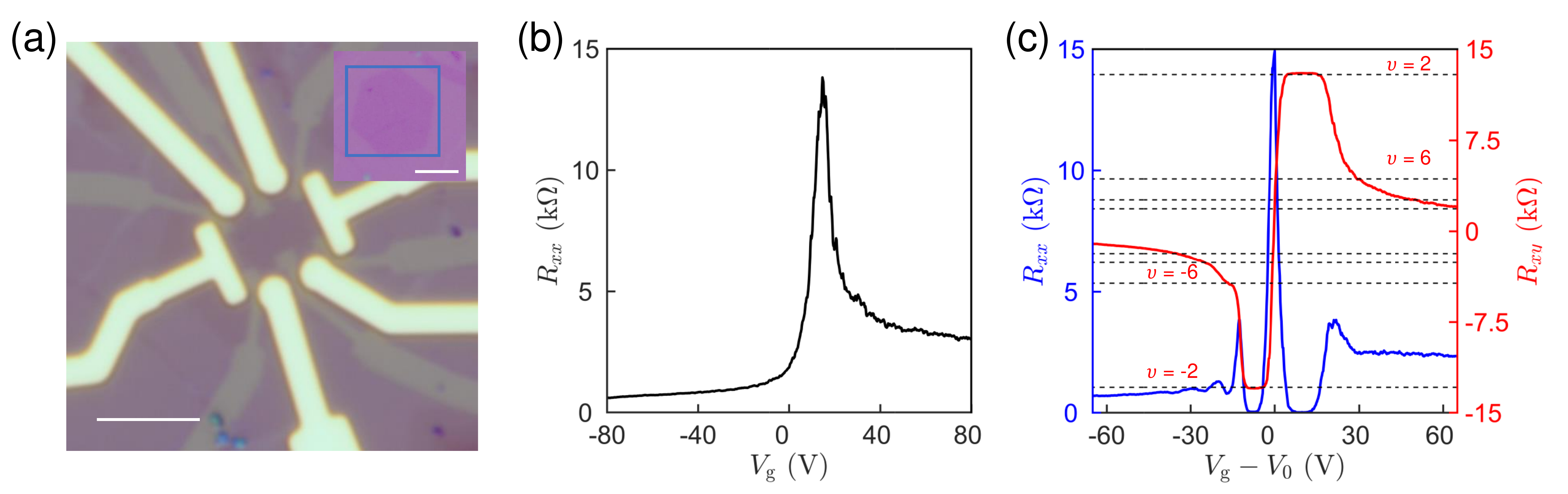}
\end{figure}
\textbf{Fig 4. Electrical Transport Measurements on Monolayer Graphene Grown from Blood Precursor. } (a) Optical images of the Hall bar device on a 300 nm $\mathrm{SiO_2/Si}$ substrate. Inset: Optical images of the same monolayer graphene grown from blood before fabrication. Scale bars, 10 $\mathrm{\mu m}$. (b) Four-probe resistance of graphene plotted against back-gate voltage at $T$ = 1.6 K. The Dirac point is clearly visible at $V_g$ = 15 V. (c) The monolayer graphene device's electrical properties display the half-integer quantum Hall effect at B = 4.5 T and $T$ = 1.6 K. The longitudinal resistance $R_{xx}$ (in blue) and transverse $R_{xy}$ (in red) are presented as a function of gate voltage. Horizontal dashed lines for various filling factors indicate quantum Hall plateaus.

\newpage
\textbf{TOC Graphic} 
\graphicspath{{Figures}}
\begin{figure}[H] 
	\centering
	\includegraphics[width=8.25cm]{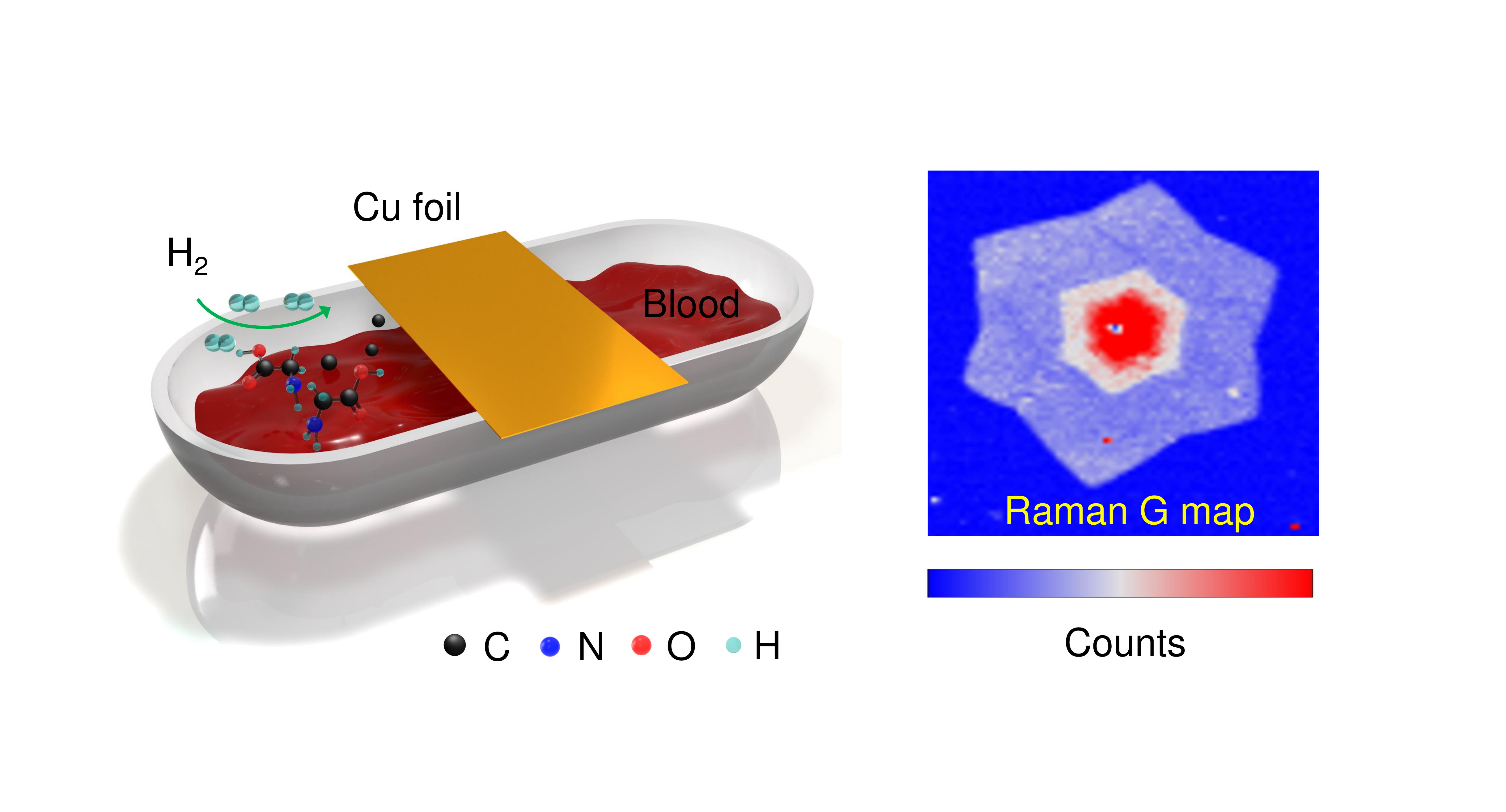}
\end{figure}

\end{document}